\documentclass[aps,prl,twocolumn,amsmath,amssymb,superscriptaddress]{revtex4-2}

\usepackage{graphicx}
\usepackage{orcidlink}
\usepackage[caption=false]{subfig}
\usepackage{dcolumn}
\usepackage{bm}
\usepackage{braket}
\newcommand{\instituteA}{Department of Physics - University of Milano-Bicocca - Piazza della Scienza 3
20126 Milan - Italy}
\newcommand{\instituteB}{INFN - Milano Bicocca - Piazza della Scienza - 3 20126 Milan - Italy}
\newcommand{\instituteC}{Bicocca Quantum Technologies (BiQuTe) Centre - Piazza della Scienza - 3 20126 Milan - Italy}

\begin{document}

\title{Achieving speedup in Dark Matter search experiments with a transmon-based NISQ algorithm}

\author{Roberto Moretti $^*$ \orcidlink{0000-0002-5201-5920}}
\affiliation{\instituteA}
\affiliation{\instituteB}
\author{Pietro Campana \orcidlink{0009-0004-8271-242X}}
\affiliation{\instituteA}
\affiliation{\instituteB}
\author{Rodolfo Carobene \orcidlink{0000-0002-0579-3017}}
\affiliation{\instituteA}
\affiliation{\instituteB}
\author{Alessandro Cattaneo \orcidlink{0009-0003-4569-1474}}
\affiliation{\instituteA}
\affiliation{\instituteB}
\author{Marco Gobbo \orcidlink{0000-0001-5543-9190}}
\affiliation{\instituteA}
\affiliation{\instituteB}
\author{Danilo Labranca \orcidlink{0000-0002-5351-0034}}
\affiliation{\instituteA}
\affiliation{\instituteB}
\author{Matteo Borghesi \orcidlink{0000-0001-5854-8894}}
\affiliation{\instituteA}
\affiliation{\instituteB}
\affiliation{\instituteC}
\author{Marco Faverzani \orcidlink{0000-0001-8119-2953}}
\affiliation{\instituteA}
\affiliation{\instituteB}
\affiliation{\instituteC}
\author{Elena Ferri \orcidlink{0000-0003-1425-3669}}
\affiliation{\instituteB}
\author{Sara Gamba \orcidlink{0009-0004-3616-7942}}
\affiliation{\instituteA}
\affiliation{\instituteB}
\author{Angelo Nucciotti \orcidlink{0000-0002-8458-1556}}
\affiliation{\instituteA}
\affiliation{\instituteB}
\affiliation{\instituteC}
\author{Andrea Giachero $^\dagger$ \orcidlink{0000-0003-0493-695X}}
\affiliation{\instituteA}
\affiliation{\instituteB}
\affiliation{\instituteC}
\begin{abstract}
Coherent detection of ultralight bosonic dark matter can be achieved by monitoring slow Rabi oscillations in superconducting qubits. We introduce an ancilla-assisted, gate-based protocol that enhances sensitivity to the hidden photon kinetic mixing parameter $\epsilon$ using a single two-qubit gate, bypassing the need to maintain long-lived multi-qubit entangled states and remaining compatible with the limitations of modern quantum hardware. We characterized the increase in sensitivity accounting for decoherence, thermal occupation, errors in readout and reset, indicating up to a ten-fold reduction in the required integration time to reach the same exclusion limit on $\epsilon$ achievable via Rabi-sampling experiments. Under plausible hardware assumptions and three years of data taking, the projected $95\%$ C.L. exclusion limit on the hidden photon mixing parameter reaches $\epsilon\approx 1\times 10^{-14}$ across $2.5$-$6.0$ GHz ($10$-$25$ \textmu eV).
\end{abstract}
\maketitle
\noindent $^*$ roberto.moretti@mib.infn.it\\
$^\dagger$ andrea.giachero@mib.infn.it

Superconducting transmon qubits are highly sensitive probes of weak electromagnetic fields in the microwave band \cite{Blais2021}. Their compatibility with cavity-based experiments makes them a versatile and powerful addition to the domain of quantum sensing. In particular, transmon-enhanced experiments have been proposed and carried out to search for ultralight bosonic dark matter (hidden photons, axions, and axion-like particles) that couple to the Standard Model electromagnetic sector \cite{kimball_vanbibber_2023}. Transmons have been incorporated into haloscopes and cavity-photon-counter experiments to reduce noise \cite{Braggio2025}, even beyond the standard quantum limit (SQL) \cite{Dixit2021, Agrawal2024, Moretti2025}, greatly enhancing sensitivity to the coupling strength parameters governing photon-dark matter conversions \cite{Browman1974,Nelson2011, Bauer2018}.

An alternative dark matter search approach consists of letting a transmon interact directly with anomalous electromagnetic signals originating from dark matter \cite{NajeraSantos2024, Kang2025}, driving its quantum state. Under the hypothesis that cold, wave-like, ultralight bosonic dark matter constitutes the local Galactic halo \cite{Blumenthal1984,PaolaArias2012}, its coupling to the ordinary photon generates a narrow-band, approximately isotropic electromagnetic field at a frequency set by the dark particle mass, $\omega_X \simeq m_X$ ($\hbar=c=1$) \cite{Chaudhuri2015}. The coherence time of this signal is set by the halo velocity dispersion, $v_\text{DM}\sim 10^{-3}$ \cite{Centers2021}. For signals in the transmon band, $\tau_\text{DM}\sim \mathcal{O}(100)$~\textmu s, similar to the coherence times of state-of-the-art transmon devices \cite{Kono2024, Tuokkola2025, Bland2025}.

A superconducting qubit with transition frequency $f_q$ immersed in the dark matter field experiences a weak Rabi drive when the qubit is tuned near resonance with dark matter ($2\pi f_q \simeq m_X$).
Preparing the qubit in the ground state and then measuring it after an interaction time $\tau$ yields, in the small coupling limit \cite{takeo1},
\begin{equation}
\begin{aligned}
    P_e(\tau) &\approx 0.04k^2\left(\frac{f_q}{1\,\text{GHz}}\right) \left(\frac{C_\Sigma}{0.1\,\text{pF}}\right)  \left(\frac{\rho_\text{DM}}{0.45\,\text{GeV/cm${}^3$}}\right) \\ &\times
    \left(\frac{d}{100\,\text{\textmu m}}\right)^2 \left(\frac{\epsilon}{10^{-11}}\right)^2 \left(\frac{\tau}{100\,\text{\textmu s}}\right)^2,
    \label{eq:darkphotonprob}
\end{aligned}
\end{equation}
where $k$ is a package–geometry coefficient (see the Supplemental Materials \cite{SM} for details), $f_q$ the qubit frequency, $C_\Sigma$ the effective transmon capacitance, $\rho_{\rm DM}$ the local dark-matter density, $d$ the qubit dipole length and $\epsilon$ the hidden photon kinetic mixing parameter. The goal of the experiment is to sample the excitation probability of a frequency-tunable transmon, as a function of $f_q$, searching for the resonance condition with the dark matter signal. An excess of excited-state counts at a specific frequency $f_X$, above expected backgrounds and statistical fluctuations, would be a candidate dark-matter signature with $m_X=2\pi f_X$.\\
An expression analogous to Eq. \eqref{eq:darkphotonprob} has also been derived for axion-induced couplings \cite{Chen2024}. Implementing such a search, however, would require operating the qubit in the presence of a few-Tesla magnetic field, decreasing the qubit's coherence. Although ongoing technical developments may eventually mitigate this limitation \cite{Schneider2019, Krause2022, Abdisatarov}, the experiment considered here focuses on hidden photon search while remaining, in principle, applicable to axion search.

\begin{figure*}
    \centering
    \includegraphics[width=0.9\linewidth]{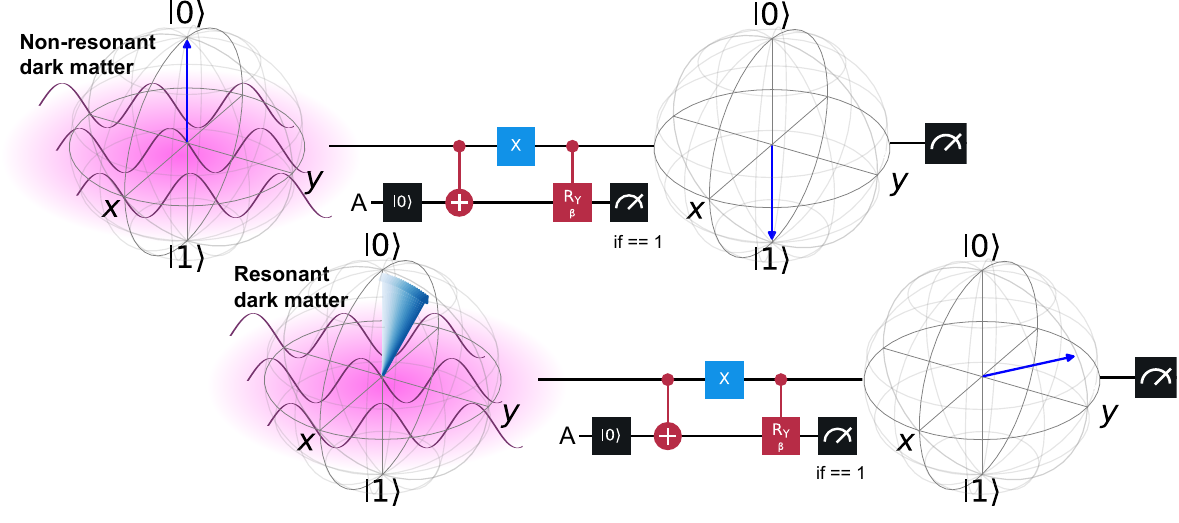}
    \caption{Enhanced direct dark matter search outline. A sensing qubit is prepared in $\ket{0_\text{S}}$ and exposed to the dark matter-induced electric field for a time $\tau$. The sensing qubit state evolution depends on whether the qubit is resonant or off-resonance with the field. Applying the enhancement circuit (requiring an ancilla qubit A) further separates the on-resonance and off-resonance states, improving state discrimination.}
    \label{fig:summary}
\end{figure*}

To increase detection sensitivity through direct qubit excitation, multi-qubit states may be leveraged. For instance, preparing $n$ qubits in a GHZ-like state yields a remarkable enhancement of the single-shot signal ($P_e\rightarrow n^2P_e$) \cite{Chen2024-2, fermilab_study} on ideal devices. In practice, multi-qubit entangled states are harder to preserve, with coherence time roughly decreasing as $1/n$ \cite{Degen2017, Kam2024}. Moreover, such enhanced detection requires all qubits to be resonant simultaneously for the duration $\tau$, which is a demanding detector engineering and operational constraint. Consequently, the $n^2$ gain is largely offset, so that $n$ independent single-qubit sensors may perform comparably well. This motivates the search for alternative multi-qubit strategies that provide a practical sensitivity advantage while remaining compatible with current hardware in the noisy intermediate-scale quantum (NISQ) era \cite{Jiang2025}.

In this Letter, we propose and analyze a NISQ era-compatible detection protocol for direct dark matter search, which yields increased sensitivity to dark matter with respect to single-qubit searches. The protocol discussed here involves one sensing qubit (S) coupled to an ancilla (A). As discussed in later paragraphs, our protocol scales to larger architectures, featuring multiple sensing qubits coupled to the same ancilla, further increasing sensitivity.\\
The core of our enhancement protocol is a transformation of the qubit state $\ket{\Psi_\tau}$ after interacting with the dark matter signal for a duration $\tau$, so that fewer measurements are required to achieve the same exclusion limit on the mixing strength parameter $\epsilon$, thus accelerating the search. In a real experiment, noise produces spurious excitation counts even when the qubit and the dark matter signal are off-resonance. We formulate the detection problem in terms of the null hypothesis $H_0$ (absence of dark matter) and the signal hypothesis $H_1$ (dark matter exists with strength $\epsilon$). The transformation $\mathcal{T}$ acts on the S-A system such that states $\mathcal{T} \ket{\Psi_\tau(H_0)}$ and $\mathcal{T}\ket{\Psi_\tau(H_1)}$ are more separated than the original $\ket{\Psi_\tau(H_0)}$ and $\ket{\Psi_\tau(H_1)}$. Fig.~\ref{fig:summary} represents our enhanced experiment workflow, while the quantum circuit implementing the transformation $\mathcal{T}$ is fully described in the Supplemental Materials \cite{SM}.\\
In the first step of the enhanced protocol, we initialize S in $\ket{0_\text{S}}$, then we let it interact coherently with the dark matter signal for a duration $\tau$. Meanwhile, A remains uncoupled from S. Right after the interaction time $\tau$, A is initialized to its ground state $\ket{0_\text{A}}$. The two-qubit quantum state reads
\begin{equation}
    \ket{\Psi_\tau} = \cos\left(\frac{\theta}{2}\right)\ket{0_\text{A}0_\text{S}} + \sin\left(\frac{\theta}{2}\right)e^{i\varphi}\ket{0_\text{A}1_\text{S}}.
    \label{eq:psi_tau}
\end{equation}
In Eq. \ref{eq:psi_tau}, we encode the Rabi oscillation of the sensing qubit into a rotation angle $\theta$ around the $y$-axis of its Bloch sphere, and into an unknown phase $\varphi$.\\
In the second step, we perform a unitary two-qubit operation, followed by a measurement of the ancilla qubit. If the ancilla collapses in $\ket{1_\text{A}}$, the two-qubit state becomes
\begin{equation}
    \ket{\Psi_\text{s}} = \frac{\sin(\frac{\beta}{2})\cos(\frac{\theta}{2})\ket{1_\text{A}1_\text{S}} + \sin(\frac{\theta}{2})e^{i\varphi}\ket{1_\text{A}0_\text{S}}}{\sqrt{\sin(\frac{\beta}{2})^2\cos(\frac{\theta}{2})^2 + \sin(\frac{\theta}{2})^2}},
    \label{eq:2qb_psi3}
\end{equation}
where $\beta$ is a free parameter of the enhancement protocol.
Upon measuring the sensing qubit, its state collapses into $\ket{0_\text{S}}$ with a probability
\begin{equation}
    \tilde{P} = \frac{P_e}{\sin(\frac{\beta}{2})^2(1-P_e) + P_e},
    \label{eq:prob_ideal}
\end{equation}
where we used the identity $P_e = \sin(\theta/2)^2$ corresponding to the excitation probability of S before enhancement. We note that the presence of the ancilla and its measurement are necessary first to extend the computational Hilbert space and then to partially collapse the two-qubit state into one that depends nontrivially on $P_e$. 
\begin{figure}
    \centering
    \includegraphics[width=\linewidth]{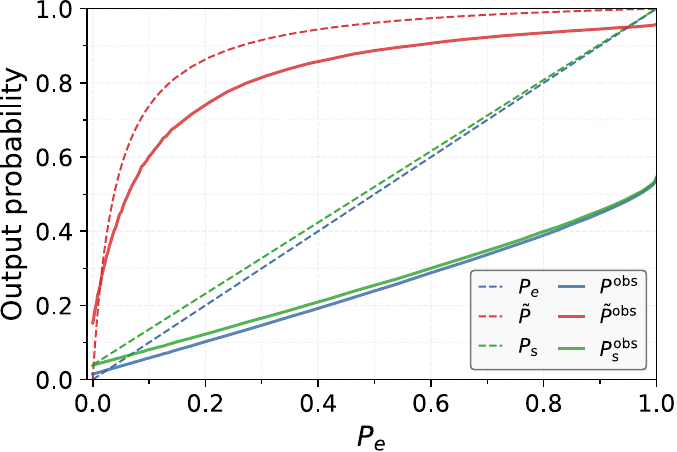}
    \caption{Excitation probability for the baseline experiment $P^\mathrm{obs}$, the conditioned enhanced signal probability $\tilde{P}^\mathrm{obs}$ (given protocol success), and the enhancement success probability $P_s^\mathrm{obs}$, plotted as a function of the ideal sensing probability $P_e$. Results correspond to the example noise configuration $\beta = 0.4$, $T_1=100$~\textmu s, $r=p=0.01$. Dashed lines show the corresponding ideal (noise-free) probabilities.}
    \label{fig:prob_curves}
\end{figure}
The slope of the $\tilde{P}$ curve near $P_e = 0$ is $\approx\sin(\beta/2)^{-2}$, i.e.\ it quickly increases as $\beta$ approaches zero. As we expect $P_e$ to be small and $P_e(H_1) > P_e(H_0)$, the protocol is advantageous because we can arbitrarily increase the response in $\tilde{P}$ for small changes in $P_e$, by tuning $\beta$.\\
This effective signal amplification only occurs when A collapses in $\ket{1_\text{A}}$ upon measurement. Otherwise, the resulting two-qubit state becomes $\ket{\Psi_\text{f}}=\ket{0_\text{A}1_\text{S}}$, carrying no information on $P_e$. We define the ancilla collapsing in $\ket{1_\text{A}}$ as a protocol \emph{success}, while the opposite case as \emph{fail}.\\
The success probability of the enhancement scheme is
\begin{equation}
\begin{aligned}
P_s &= \sin\left(\frac{\beta}{2}\right)^2\cos\left(\frac{\theta}{2}\right)^2 + \sin\left(\frac{\theta}{2}\right)^2 \\
&= \sin\left(\frac{\beta}{2}\right)^2 + \cos\left(\frac{\beta}{2}\right)^2 P_e,
\end{aligned}
\label{eq:psuccess}
\end{equation}
hence near $P_e=0$ it decreases as $\beta$ decreases. Thus, the choice of $\beta$ trades off the amount of signal amplification in case of success and the success probability itself. The optimal choice of $\beta$ ultimately depends on the noise specifications of the quantum device in use.\\
The enhancement circuit requires only one two-qubit entangling gate when transpiled to native gate sets common to most quantum processing units: echoed cross resonance or controlled-Z. The protocol can therefore be implemented on current NISQ devices as they nowadays exhibit two-qubit gate fidelities well above $99\%$ \cite{Li2024, marxer2025}.

We quantified the protocol’s advantage relative to a baseline single-qubit direct-detection experiment under realistic, NISQ-like noise. A gate-based noise model was built, including energy relaxation ($T_1$) and pure dephasing ($T_\phi$) times of both the sensing and ancilla qubits during $\tau$ and the protocol operations (gates, state reset, and measurement). Effective qubit temperature $T_\text{eff}$ \cite{Krantz2019, Lvov2025}, state-preparation error rates $p$, and readout assignment error rates $r$ were also included. Details on the noise channel implementations and comparison between simulations and data computed on IBM real quantum backends are reported in the Supplemental Materials \cite{SM}. We emphasize how our enhancement protocol can be tested directly on cloud-access quantum hardware, yielding results that are consistent with our simulations.

\begin{figure}
    \centering
    \includegraphics[width=0.95\linewidth]{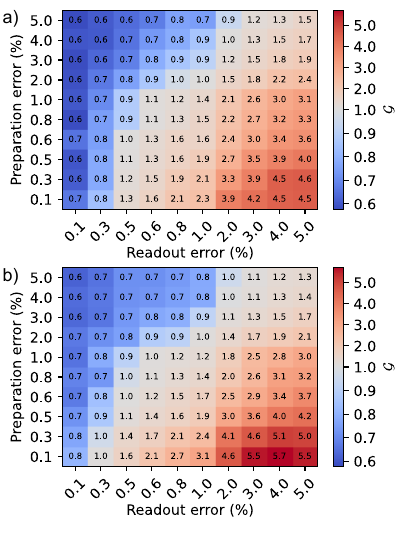}
    \caption{Speedup factor $\mathcal{G}$ of the two-qubit enhanced protocol relative to two parallel single-qubit baseline experiments, computed for $m_X=4.5$~GHz. The speedups are evaluated on an $r$-$p$ grid with both errors varying from $0.1\%$ to $5.0\%$. (a) Medium coherence regime, with $Q=\pi\times 10^6$ for both sensing and ancilla. (b) High coherence scenario, with $Q=2\pi\times 10^6$ for both sensing and ancilla.}
    \label{fig:speedups}
\end{figure}

The experiment requires the sensing qubit to be tunable, but it also highly benefits from a high dephasing time, even when biased far from its sweetspots. For this reason, we constrain the dynamic range to $300$ MHz, so that the decrease in $T_\phi$ far from the sweetspots is tolerable \cite{Hutchings2017, Vepslinen2022, ChvezGarcia2022}. To span a larger frequency interval, more sensing qubits must be designed to cover the desired spectrum, either in separate detectors or within the same chip.\\
In our simulation, the relaxation times are treated as frequency-dependent by fixing the quality factor $Q$. We examined two regimes, referred to as \emph{medium-coherence} ($Q = \pi \times 10^6$) and \emph{high-coherence} ($Q = 2\pi \times 10^6$), corresponding to $T_1=100$~\textmu s and $T_1=200$~\textmu s at $f_q=5$ GHz for both S and A. To account for a realistic dephasing time, we set $T_\phi=2T_1$ (hence $T_2=T_1$). We considered an effective qubit temperature $T_\mathrm{eff}=35$ mK, in agreement with residual thermal excitations measured in the literature for standard setups \cite{Jin2015, Kulikov2020, Sun2023}.
The noise simulation returns the observed excitation $P_\mathrm{base}^\mathrm{obs}$ in a baseline experiment, the enhanced excitation $1-\tilde P^\mathrm{obs}$ (given the protocol's success), and the enhancement success $P_s^\mathrm{obs}$. Each of these observables depends on the ideal excitation probability prior to enhancement $P_e$, which depends on $\epsilon$, and on the sensing qubit probe frequency $f_q$. Evaluating the noise model over a two-dimensional grid in ($f_q,P_e$) thus defines the mapping
\begin{equation}
\mathcal{M}(f,P_e;\mathbf{k})=\begin{pmatrix}P^\mathrm{obs},\ \tilde P^\mathrm{obs},\ P_s^\mathrm{obs}\end{pmatrix},
\end{equation}
which we use to generate toy experiments and quantify the baseline and enhanced experiment sensitivities to $\epsilon$. The parameter vector $\mathbf{k}$ collects the experimental and noise model inputs \{$Q$, $\tau$, $T_\text{eff}$, $p$, $r$, $\beta$\}. An example outcome is shown in Fig.~\ref{fig:prob_curves}, comparing the ideal $P_e$, $\tilde{P}$, $P_s$ with their noisy counterparts $P^\mathrm{obs}_e$, $\tilde{P}^\mathrm{obs}$, $P^\mathrm{obs}_s$.\\
To extract an exclusion limit on $\epsilon$, we perform a hypothesis test between the signal-background hypotheses $H_0$ and $H_1$ using a binned log-likelihood-ratio statistic \cite{1933}
\begin{equation}
q = -2\left[\ln\mathcal{L}(H_0) - \ln\mathcal{L}(H_1)\right],
\end{equation}
where $\mathcal{L}(H_0)$ and $\mathcal{L}(H_1)$ denote the likelihood evaluated on arrays of observed counts across probe frequency bins. Details on the likelihood construction, including a method for incorporating additional information from failed enhancement attempts, are given in the Supplemental Materials \cite{SM}.\\
For each value of $\epsilon$ in input, we compute the statistic $q$ and obtain the corresponding $p$-value for the null hypothesis. The $95\%$ confidence-level (C.L.) exclusion on $\epsilon$ is then defined as the value of $\epsilon$ that yields $p\text{-value}=0.05$.\\
We explored a two-dimensional grid in the dominant error parameters $r$ and $p$, while keeping the other noise and experimental parameters fixed, determining the $95\%$ C.L. base and enhanced limits $\epsilon^\text{base}_{95\%}$ and $\epsilon^\text{enh}_{95\%}$. Under the counting-statistics–dominated assumption, the minimum detectable coupling strength scales with the total integration time $T$ as $\epsilon \propto T^{-1/4}$, as a consequence of $P_e\propto \epsilon^2$ and the counting experiment standard deviation scaling as $\propto T^{-1/2}$. Hence we define the speedup factor
\begin{equation}
    \mathcal{G}  = \left(\frac{\epsilon^\text{base}_{95\%}}{\epsilon^\text{enh}_{95\%}}\right)^4,
\end{equation}
computed using $10^6$ measurements per probe frequency for the enhanced experiment and twice as many for the baseline one: because the enhanced protocol uses two qubits, we compare it with two identical single-qubit experiments running in parallel. We note that this correction does not account for potential overheads such as additional calibration or control complexity in the enhancement protocol.
Fig.~\ref{fig:speedups} summarizes the speedup factors computed on the $r$-$p$ grid for a hidden photon mass of $m_X=4.5$~GHz. Panels (a) and (b) show the medium and high coherence scenarios, respectively. In both cases, we fixed $\tau = 100$~\textmu s, $T_\mathrm{eff}=35$~mK, and $\beta=0.2$.
\begin{figure*}[ht]
    \centering
    \includegraphics[width=0.9\linewidth]{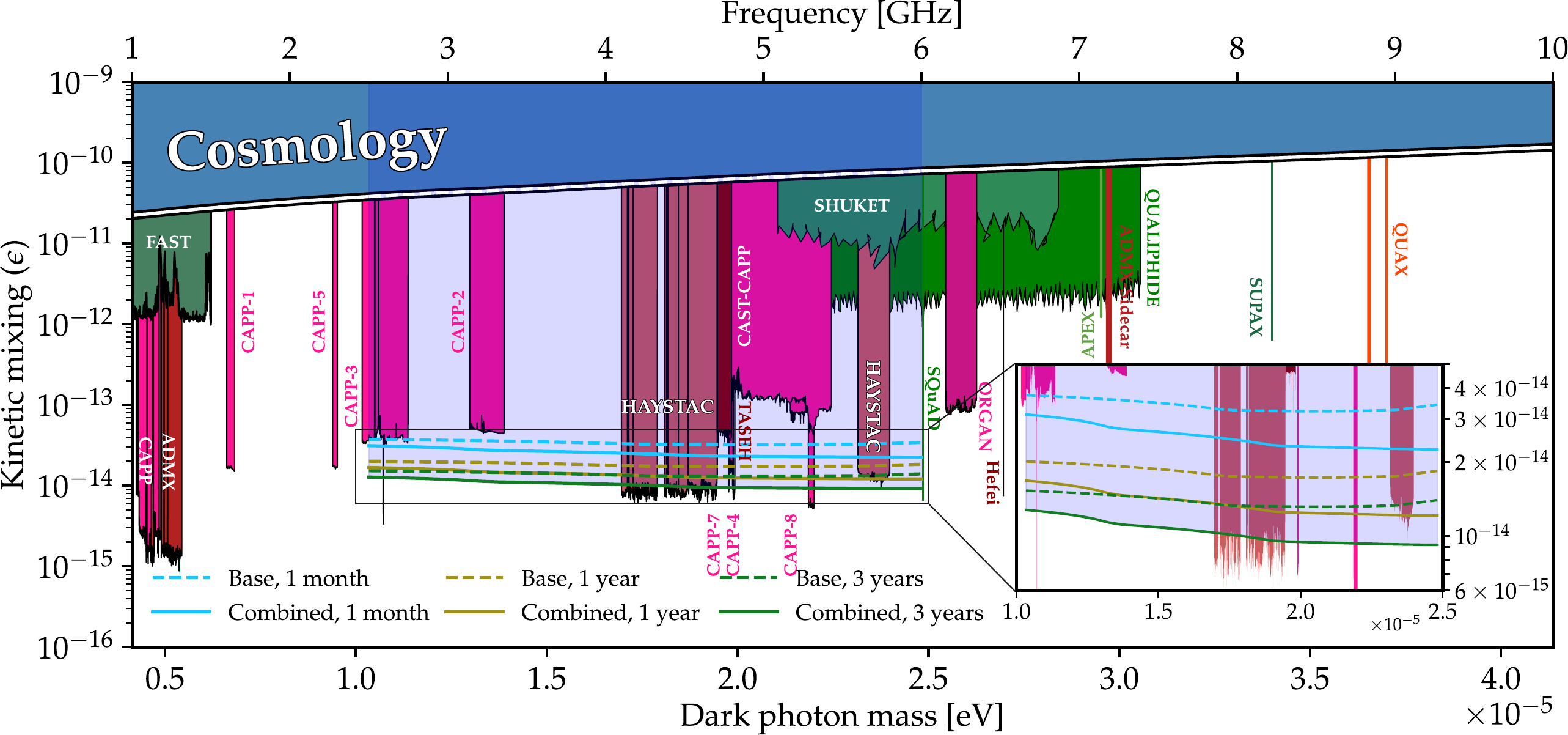}
    \caption{Smooth envelope of the projected 95\% C.L. exclusion limits on the hidden-photon kinetic-mixing parameter $\epsilon$ for the baseline and enhanced protocols, on top of existing haloscope limits and cosmological bounds in the $1$-$10$~GHz band \cite{AxionLimits}. }
    \label{fig:limitprojection}
\end{figure*}
The simulation reveals two qualitatively distinct regimes. When the readout error $r$ exceeds the state preparation error $p$, the enhanced experiment yields a speedup factor above one in almost all cases. The largest speedups ($\mathcal{G}\sim 4$-$6$) occur for a preparation error $p\lesssim 0.3\%$, combined with a readout error $r\gtrsim 2\%$. On the other hand, the enhanced experiment can perform worse than two parallel baseline experiments ($\mathcal{G}<1$) when $r<0.3\%$, or $p>2\%$.\\
We note that active reset protocols often suppress the residual excited-state population to the order of $0.1\%$ \cite{Zhou2021, Kim2025, Magnard2018}, whereas achieving single-shot readout error reliably below $0.3\%$, despite being demonstrated \cite{Spring2025}, remains challenging and often requires methods beyond standard dispersive readout \cite{lowr1, Hazra2025, Gao2025}, making the enhancement protocol advantageous in many practical scenarios.
The difference in $\mathcal{G}$ between the medium and high coherence regimes is modest, as the same qualitative trends appear in both cases. Between $2.5$ GHz and $6.0$ GHz (corresponding to the mass interval $10$-$25$ \textmu eV), $\mathcal{G}$ increases with $m_X$ (see Supplemental Materials \cite{SM}). We attribute this effect to the frequency dependence of the thermal background, with higher probe frequencies being less susceptible to spurious thermal excitations, which are amplified by the enhancement protocol.

The speedup factors reported in Fig.~\ref{fig:speedups} refer to the two-qubit case. Further speedup is accessible by coupling more sensing qubits to the same ancilla. In fact, the interaction time $\tau \sim \min(\tau_\mathrm{DM}, T_1)\sim 100$~\textmu s, is much longer than the enhancement protocol duration, determined by a few $\mathcal{O}(1)$~\textmu s operations. With a fast ancilla reset, the same ancilla can be reused to enhance multiple sensing qubits sequentially. For example, two sensing qubits can co-evolve for a time $\tau$, then the enhancement circuit is applied between one sensing qubit and the ancilla, the ancilla is then reset, and undergoes the enhancement circuit with the second sensing qubit. This yields two enhanced samples within the same interrogation window $\tau$, while adding only one physical qubit.\\
Extending this principle to an $n$-qubit setup yields an effective speedup
\begin{equation}
    \mathcal{G} \rightarrow 2\mathcal{G}\frac{n-1}{n}
\end{equation}
over an $n$-qubit base experiment counterpart. The overall speedup factor then approaches $2\mathcal{G}$ as many sensing qubits are coupled to the same ancilla, increasing the speedup to $\mathcal{G}\sim 10$ in the low-$p$, high-$r$ scenario.

In the limit where many sensing qubits are coupled to the same ancilla ($\mathcal{G}\rightarrow 2\mathcal{G}$), we carried out a sensitivity projection for an experiment comprising a total of $120$ physical qubits, deployed either within a single module or across multiple modules operating in parallel. With a dynamic range of $300$ MHz per sensing qubit, the $2.5$-$6.0$ GHz band is partitioned into $12$ frequency sub-intervals, each assigned $10$ qubits. Keeping $\tau = 100$~\textmu s, we add a $30$~\textmu s overhead per sample to account for qubit operations and classical data registration. Probe frequencies are sampled on an evenly spaced grid with a step size of $100$~kHz. With these experimental parameters we estimate to collect around $80\times 10^6$ measurements per year at each frequency.
Fig.~\ref{fig:limitprojection} presents the projected exclusion limits on $\epsilon$ for a dipole length $d=250$~\textmu s, computed in the high-coherence regime with error rates $r=5\times10^{-3}$ and $p=10^{-3}$. After three years of active data taking, the projected exclusion boundary reaches $\epsilon_\text{$95\%$ C.L.} \approx1\times 10^{-14}$ across the probed frequency interval. Compared to haloscope searches, the proposed experiment spans a substantially larger frequency range, despite remaining intrinsically narrow-band and constrained by probing discrete probe-frequency steps.
In conclusion, we introduced a gate-based protocol enhancing the experimental sensitivity to resonant dark-matter drives without requiring long-lived, large-scale multi-qubit entanglement during the sensing interval. Under NISQ-like noise, the protocol yields a tangible sensitivity improvement when readout assignment errors are larger than state preparation errors. The approach is extendable to axion searches if qubit architectures that tolerate strong magnetic fields become available. A near-term development would consist of an experimental demonstration of an enhanced dark matter search, enabling precise quantification of systematic uncertainties.

\textit{Acknowledgements}--This work is supported by QUART\&T, a project funded by the Italian Institute of Nuclear Physics (INFN) within the Technological and Interdisciplinary Research Commission (CSN5) and Theoretical Physics Commission (CSN4) and by  PNRR MUR projects PE0000023-NQSTI and CN00000013-ICSC. The authors acknowledge their valuable discussions with Michele Grossi and Mikio Nakahara.

\textit{Data Availability}--The data that support the findings of this article are openly available \cite{qubitm}.

\bibliography{refs}   

\end{document}


\begin{center}
    \Large{SUPPLEMENTAL MATERIAL}
\end{center}
\title{Achieving speedup in Dark Matter search experiments with a transmon-based NISQ algorithm}

\author{Roberto Moretti $^*$ \orcidlink{0000-0002-5201-5920}}
\affiliation{\instituteA}
\affiliation{\instituteB}
\author{Pietro Campana \orcidlink{0009-0004-8271-242X}}
\affiliation{\instituteA}
\affiliation{\instituteB}
\author{Rodolfo Carobene \orcidlink{0000-0002-0579-3017}}
\affiliation{\instituteA}
\affiliation{\instituteB}
\author{Alessandro Cattaneo \orcidlink{0009-0003-4569-1474}}
\affiliation{\instituteA}
\affiliation{\instituteB}
\author{Marco Gobbo \orcidlink{0000-0001-5543-9190}}
\affiliation{\instituteA}
\affiliation{\instituteB}
\author{Danilo Labranca \orcidlink{0000-0002-5351-0034}}
\affiliation{\instituteA}
\affiliation{\instituteB}
\author{Matteo Borghesi \orcidlink{0000-0001-5854-8894}}
\affiliation{\instituteA}
\affiliation{\instituteB}
\affiliation{\instituteC}
\author{Marco Faverzani \orcidlink{0000-0001-8119-2953}}
\affiliation{\instituteA}
\affiliation{\instituteB}
\affiliation{\instituteC}
\author{Elena Ferri \orcidlink{0000-0003-1425-3669}}
\affiliation{\instituteB}
\author{Sara Gamba \orcidlink{0009-0004-3616-7942}}
\affiliation{\instituteA}
\affiliation{\instituteB}
\author{Angelo Nucciotti \orcidlink{0000-0002-8458-1556}}
\affiliation{\instituteA}
\affiliation{\instituteB}
\affiliation{\instituteC}
\author{Andrea Giachero $^\dagger$ \orcidlink{0000-0003-0493-695X}}
\affiliation{\instituteA}
\affiliation{\instituteB}
\affiliation{\instituteC}
\maketitle
\noindent $^*$ roberto.moretti@mib.infn.it\\
$^\dagger$ andrea.giachero@mib.infn.it
\section{Derivation of the enhancement protocol}
In natural units ($c=\hbar = 1$,), the coupling of hidden photons to the electron and the ordinary photon is described by the interaction term
\begin{equation}
    \mathcal{L}_\text{int} = e\bar{\psi}_e\gamma^\mu\left(A_\mu + \epsilon X_\mu\right)\psi_e,
\end{equation}
where $\psi_e$, $A_\mu$, and $X_\mu$ are the electron, photon, and dark photon quantum fields, respectively. The term $\epsilon$ is the kinetic-mixing parameter between hidden and ordinary photons. For a monochromatic dark-photon field of mass $m_X$ we write the vector potential as
\begin{equation}
    \mathbf{X} = \bar{X}\mathbf{n}_X \cos(m_X t),
\end{equation}
where $\mathbf{n}_X$ is a unit vector and $\bar{X}$ the oscillation amplitude, linked to the local DM density $\rho_\text{DM} = m_X^2\bar{X}^2/2$.
The dark component of the electric field on the qubit is, therefore,
\begin{equation}
    \mathbf{E}_X = -\epsilon \frac{\partial{\mathbf{X}}}{\partial t} = \epsilon m_X \mathbf{n}_X \bar{X} \sin(m_X t).
\end{equation}
Interactions of the induced field with the metallic package walls (readout cavity for 3D qubit architectures, or sample holder for 2D architectures) produce a reactive field $\mathbf{E}_{\rm EM}$~\cite{Huang2021} such that the net effective field inside the package is
\begin{equation}
    \mathbf{E}_{\rm eff}(t)=\mathbf{E}_X(t)+\mathbf{E}_{\rm EM}(t).
\end{equation}
We define a package factor \(k\equiv \bar{E}_{\rm eff}/\bar{E}_X\), and, using \(\rho_{\rm DM}=m_X^2\bar{X}^2/2\), we obtain
\begin{equation}
    \mathbf{E}_{\text{eff}} = \bar{E}_{\text{eff}} \mathbf{n}_{\text{eff}} \sin(m_X t) = k \epsilon \sqrt{2\rho_{\text{DM}}}\mathbf{n}_{\text{eff}} \sin(m_X t).
    \label{eq:darkphoton_eff_field}
\end{equation}
Following the derivation presented in \cite{takeo1}, when the transmon qubit is resonant with the dark matter signal ($m_X=\omega_q$), the driven two-level dynamics under the rotating-wave approximation (RWA) reduce to
\begin{equation}
    i\frac{d}{dt}\begin{pmatrix}
        \psi_g(t) \\
        \psi_e(t)
    \end{pmatrix} \approx \begin{pmatrix}
        0 & -i\eta \\
        i\eta & 0
    \end{pmatrix} \begin{pmatrix}
        \psi_g(t) \\
        \psi_e(t)
    \end{pmatrix},
    \label{eq:slowrabidyn2}
\end{equation}
with
\begin{equation}
    \eta \equiv \frac{1}{2}\sqrt{\frac{\omega_q C_\Sigma} {2}} d \bar{E}_\text{eff} \cos(\Theta) = \frac{\sqrt{\omega_q C_\Sigma\rho_\text{DM}}}{2}d k \epsilon \cos(\Theta),
    \label{eq:eta_formula}
\end{equation}
where $d$ is the qubit dipole length, $C_\Sigma$ its effective capacitance, and $\Theta$ the angle between the dipole orientation and $\mathbf{E}_\text{eff}$.

Considering a qubit initially prepared in $\ket{0}$, hence $\psi_g(0) = 1$ and $\psi_e(0) = 0$, the solutions to Eq.~\ref{eq:slowrabidyn2} are:
\begin{equation}
    \psi'_g(t) = \cos(\eta t); \,\,\,\,\,\,\,\, \psi'_e(t) = \sin(\eta t).
\end{equation}
The probability of the qubit to collapse in $\ket{1}$ after a measurement is
\begin{equation}
    P_e(t) = |\psi'_e(t)|^2 = \sin^2\left(\frac{\sqrt{\omega_q C_\Sigma\rho_\text{DM}}}{2}d k \epsilon \cos(\Theta) t\right).
\end{equation}
We expect the package-geometry coefficient to be order unity \cite{takeo1}, therefore we assume $k=1$ throughout the manuscript.
Since we expect the coupling to be small, we assume that a complete Rabi cycle would take much longer than the qubit's (or the dark matter-induced signal) coherence time. Hence, we approximate the Rabi oscillations to the second order in $t$. Under the assumption that the hidden photon-driven electric field is unpolarized \cite{Brun2019, taseh} or, equivalently, averaging over all the possible hidden photon polarizations \cite{Caputo2021} (as data-taking occurs while Earth rotates in the galactic halo), the term $\cos^2(\Theta)$ is substituted with a factor $1/3$. \\
The dark matter-driven time-evolution operator can be read as
\begin{equation}
    U_\text{DM}(t) = \begin{pmatrix}
        \cos(\eta t) & ie^{-i\varphi}\sin(\eta t) \\
        ie^{i\varphi}\sin(\eta t) & \cos(\eta t)
    \end{pmatrix}.
\end{equation}
We decompose the action of $U\text{DM}$ on the sensing qubit as a $\text{R}_\text{Y}(\theta)$ gate followed by a phase gate $\text{P}(\varphi)$. We note that $\theta \approx 2\eta t$ for $\eta t \ll 1$, while $\varphi$ is unknown \cite{Chen2024}. With a sensing qubit (S) and an ancilla (A) initialized in $\ket{0_\text{A}0_\text{S}}$, the post-drive state after an interaction time $t=\tau$ is
\begin{equation}
    \ket{\Psi_\tau} = \cos\left(\frac{\theta}{2}\right)\ket{0_\text{A}0_\text{S}} + \sin\left(\frac{\theta}{2}\right)e^{i\varphi}\ket{0_\text{A}1_\text{S}},
\end{equation}
We note that active state reset techniques can be leveraged to prepare the ancilla qubit in $\ket{0_\text{A}}$ right before the time interval $\tau$ ends, to minimize the ancilla state decoherence prior to applying the enhancement protocol.

\begin{figure}
    \centering
    \includegraphics[width=0.85\linewidth]{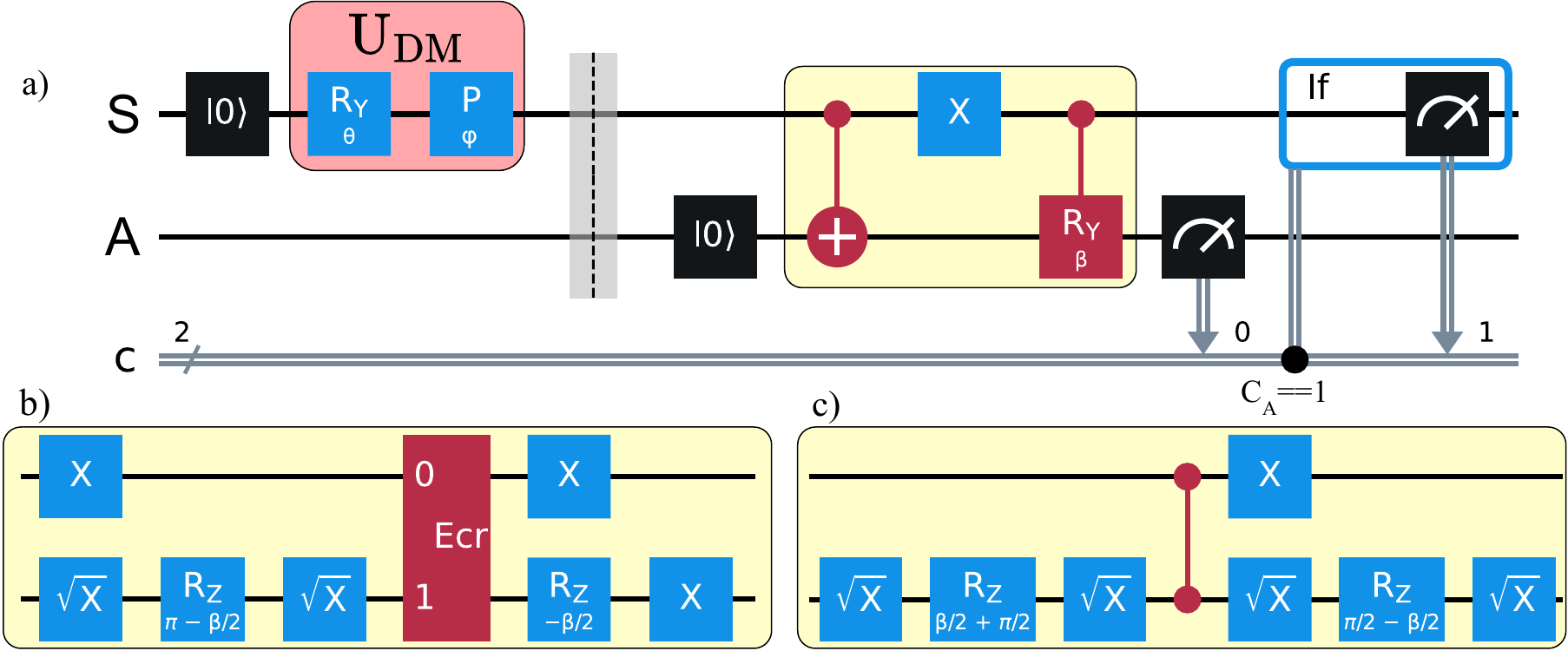}
    \caption{(a) Quantum-circuit schematic of the dark-matter sensitivity–enhancement protocol. A sensing qubit S is initialized in $\ket{0_\text{S}}$, allowed to evolve under the dark-matter drive $\text{U}_\text{DM}$ (modeled here as an $\text{R}_\text{Y}(\theta)$ followed by a phase rotation of angle $\varphi$). The qubit S then entangles with an ancilla A in the dark matter signal enhancement sequence. (b,c) Two equivalent transpiled implementations of the enhancement gate sequence expressed in different native-gate sets: (b) uses \{I, X, SX, RZ, CZ\} and (c) uses \{I, X, SX, RZ, ECR\}.}
    \label{fig:circuit}
\end{figure}
We now describe step by step how the state $\ket{\Psi_\tau}$ transforms by applying the enhancement circuit, depicted in Fig.\ref{fig:circuit}. Applying a controlled-NOT (CNOT) with S as control and A as target, followed by an X-gate on S, the two-qubit quantum state becomes
\begin{equation}
    \text{X}_\text{S}\text{CNOT}_{\text{S}\to \text{A}}\ket{\Psi_\tau}
    = \cos\left(\frac{\theta}{2}\right)\ket{0_\text{A}1_\text{S}} + \sin\left(\frac{\theta}{2}\right)e^{i\varphi}\ket{1_\text{A}0_\text{S}}.
\end{equation}
A subsequent controlled-$\text{R}_\text{Y}(\beta)$ (CRY($\beta$)) with S control and A target yields
\begin{equation}
    \text{CRY}(\beta)_{\text{S}\to \text{A}} \text{X}_\text{S}\text{CNOT}_{\text{S}\to \text{A}}\ket{\Psi_\tau} = \cos\left(\frac{\beta}{2}\right)\cos\left(\frac{\theta}{2}\right)\ket{0_\text{A}1_\text{S}}
    + \sin\left(\frac{\beta}{2}\right)\cos\left(\frac{\theta}{2}\right)\ket{1_\text{A}1_\text{S}}  + \sin\left(\frac{\theta}{2}\right)e^{i\varphi}\ket{1_\text{A}0_\text{S}}.
\label{eq:state_before_measurement}
\end{equation}

Post-selecting on the ancilla outcome $\ket{1_\text{A}}$, i.e.\ the success condition for the enhancement protocol, produces
\begin{equation}
    \ket{\Psi_s} = \frac{\sin(\beta/2)\cos(\theta/2)\ket{1_\text{A}1_\text{S}} + \sin(\theta/2)e^{i\varphi}\ket{1_\text{A}0_\text{S}}}
    {\sqrt{\sin^2(\beta/2)\cos^2(\theta/2) + \sin^2(\theta/2)}}.
\end{equation}
The conditional probability to find the sensing qubit in $\ket{0_\text{S}}$ is therefore
\begin{equation}
    \tilde{P} = \frac{\sin^2(\theta/2)}{\sin^2(\beta/2)\cos^2(\theta/2) + \sin^2(\theta/2)},
\end{equation}
or, using \(P_e=\sin^2(\theta/2)\),
\begin{equation}
    \tilde{P} = \frac{P_e}{\sin^2(\beta/2)(1-P_e) + P_e},
    \label{eq:prob_ideal}
\end{equation}
which for $P_e\ll1$ yields $\tilde{P}\approx P_e\sin^{-2}(\beta/2)$, giving the signal enhancement effect. We note that the detuning $\delta$ between the qubit's probe frequency and $m_X$ modifies $\theta$, but not the protocol algebra, which extends to any detuning.

\section{Noise modeling and CPTP simulations}

\begin{figure}[t]
    \centering
    \begin{subfigure}{0.6\textwidth}
        \centering
        \includegraphics[width=\textwidth]{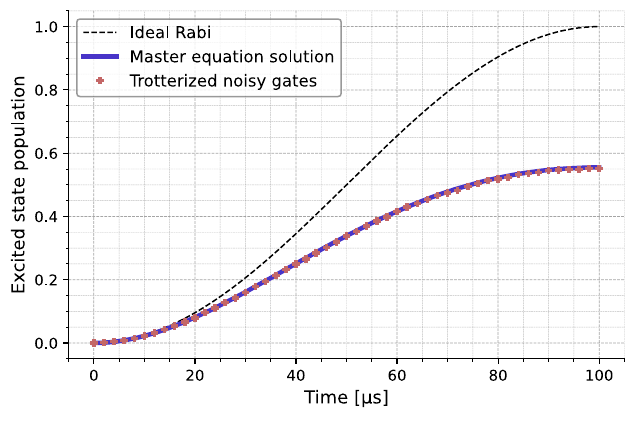}
        \caption{Comparison of Trotterized gate-model evolution with direct Lindblad integration (analytic/numerical).}
        \label{fig:check_noise}
    \end{subfigure}
    \begin{subfigure}{0.9\textwidth}
        \centering
        \includegraphics[width=\textwidth]{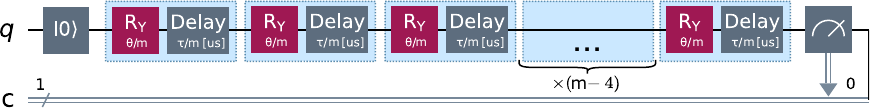}
        \caption{Quantum circuit implementing the Trotterized Rabi drive: $m$ repetitions of $\text{R}_\text{y}(\theta/m)$ each followed by a delay-CPTP channel.}
        \label{fig:trotterized_rabi}
    \end{subfigure}
    
    \caption{Accuracy of the gate-based noise modeling of a slow, resonant Rabi oscillation in the presence of relaxation and dephasing. (a) Example evolution for $T_1=T_2=100$ \textmu s ($T_\phi = 200$ \textmu s), zero thermal population, and ideal readout ($r=0$). (b) Circuit representation used in simulations.}
    \label{fig:trotter_study}
\end{figure}

We implement a fully gate-based noise model using completely positive, trace-preserving (CPTP) maps represented in Kraus form \cite{Nielsen2012}. Decoherence during the dark matter interaction interval is modeled by a Trotter-like decomposition: the slow Rabi drive is approximated by $m$ small rotations $\text{R}_\text{Y}(\theta/m)$, each followed by a CPTP map that models relaxation and dephasing acting for a time $\tau/m$. In the Lindblad picture, defining $\mathcal{L}_U[\rho]$ and $\mathcal{L}_D[\rho]$ as the non-commuting superoperators describing the unitary drive and the dissipative processes, respectively, the time derivative of the density matrix is
\begin{equation}
    \dot{\rho} = \mathcal{L}_U[\rho] + \mathcal{L}_D[\rho].
\end{equation}
The solution for $\dot{\rho}$ is found by approximating $e^{(\mathcal{L}_U+\mathcal{L}_D)\tau}$ with $\left(e^{\mathcal{L}_U\tau/m}e^{\mathcal{L}_D\tau/m}\right)^m$. In practice, the sensing qubit time evolution during the dark matter-evolution $\tau$ in the presence of decoherence is constructed by combining the ideal rotation on the Pauli basis with the Choi representation of the CPTP map. The Choi matrix $J_\Lambda$ of a single-qubit channel $\Lambda$ is diagonalized to obtain eigenvalues and eigenvectors, which are reshaped into Kraus operators. In our model we parameterize the per-step relaxation probability $p_1 = 1 - e^{-\tau/(nT_1)}$, dephasing $p_\phi = 1 - e^{-\tau/(nT_\phi)}$, and the thermal excited-state population $p_{\rm th}$ (Bose–Einstein occupation at effective temperature $T_{\rm eff}$. The resulting CPTP maps are composed with ideal gates using the standard Choi-Jamiolkowski isomorphism.\\
Fig.~\ref{fig:trotter_study} compares the Trotterized gate-based simulation with a direct numerical integration of the Lindblad master equation, showing good agreement for the chosen step number $m=50$. Practically, noise and decoherence are implemented at each gate by leveraging the \texttt{AerBackend} of the Qiskit software development kit (SDK) \cite{qiskit2024}, applying custom error channels.

\subsection{Gate set, errors, and leakage}
\begin{figure}[t]
    \centering
    \includegraphics[width=0.7\textwidth]{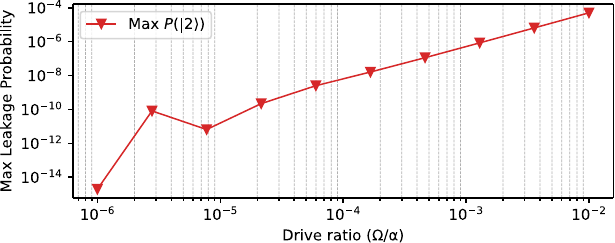}
    \caption{Estimated maximum population of state $\ket{2}$ for a qubit with anharmonicity $\alpha$ under a weak resonant drive, over a full Rabi cycle. Leakage is negligible for the parameters used in this study.}
    \label{fig:leakage_test}
\end{figure}

We embed the decoherence channels into each operation of the native gate set $\{\text{X},\ \text{SX},\ \text{I},\ \text{R}_\text{Z},\ \text{ECR}\}$, chosen because it allows the implementation of the enhancement circuit with a single two-qubit gate. Gate durations are assigned following typical superconducting-processor specifications (see Table \ref{tab:ibm_characterizations}). During each gate, the qubit undergoes relaxation and dephasing with rates set by $T_1$ and $T_\phi$. The effect of decoherence at this stage, however, is much less evident than during the interaction time $\tau$, as the gate pulse lengths are much shorter than $\tau$. \\
Measurement assignment errors were applied to the outcome probabilities via a classical confusion matrix. In our sensitivity study, for simplicity, readout errors $r$ are modeled as symmetric assignment probabilities, i.e.\ the probability of registering state $\ket{0}$ when the qubit collapses in $\ket{1}$ and vice versa are set to equal. Similarly, the probability of preparation state error $p$ is set to be equal in preparing state $\ket{0}$ and $\ket{1}$. This process was handled automatically within the Qiskit Aer-simulated backend.\\
Cooper-pair population drift toward thermal equilibrium was modeled as relaxation to the thermal steady state (with a time constant $T_1$), with level populations set by the Bose–Einstein distribution \cite{Krantz2019, Lvov2025}.\\
Leakage to the $\ket{2}$ level of the sensing qubit during the dark matter-driven evolution was estimated by solving the truncated three-level Hamiltonian
\begin{equation}
    \hat{\mathcal{H}} = \omega_q \hat{a}^\dagger \hat{a} - \frac{\alpha}{2}\hat{a}^\dagger\hat{a}^\dagger\hat{a}\hat{a}
    + \Omega(\hat{a}^\dagger + \hat{a}),
\end{equation}
with $\alpha$ the qubit's anharmonicity, and $\Omega$ the drive strength. Assuming $\Omega/\alpha \ll 10^{-5}$ (corresponding to Rabi oscillations slower than $100$ \textmu s), numerical results (Fig.~\ref{fig:leakage_test}) show leakage probabilities $<10^{-10}$, justifying the two-level approximation during the dark matter-driven evolution.

\section{Hardware demonstration on IBM Quantum backends}
To validate the noise model, we implemented the two-qubit enhancement circuit on two IBM Eagle r3 backends (\texttt{ibm\_sherbrooke} and \texttt{ibm\_brisbane}), with the native gate set $\text{X}, \text{SX}, \text{I}, \text{RZ}, \text{ECR}$. We selected the sensing-ancilla pair according to the coupling map of each device and the calibration data publicly available at the time of the experiment, summarized in Table~\ref{tab:ibm_characterizations}. We traded off ECR error rate, coherence times, state preparation error, and readout fidelities.
\begin{table}[t]
\centering
\caption{Qubit characterization used for hardware tests. Reported readout errors are conditional assignment probabilities. Gate and ECR errors are reported in the table units.}
\label{tab:ibm_characterizations}
\setlength{\tabcolsep}{8pt} 
\renewcommand{\arraystretch}{1.2} 
\begin{tabular}{lccccc}
\toprule
 & \multicolumn{2}{c}{\textbf{ibm\_sherbrooke}} &
   \multicolumn{2}{c}{\textbf{ibm\_brisbane}} \\
\cmidrule(lr){2-3} \cmidrule(lr){4-5}
Role & sensing & ancilla & sensing & ancilla \\
Qubit id & 40 & 39 & 118 & 119 \\
$T_1$ \([\mu\mathrm{s}]\)$\, $ & 251 & 361 & 151 & 159 \\
$T_2^*$ \([\mu\mathrm{s}]\)$\, $ & 52 & --- & 34 & --- \\
$T_2$ \([\mu\mathrm{s}]\)$\, $ & 327 & 281 & 126 & 69 \\
$\omega_q/2\pi$ [GHz] & 4.701 & 4.574 & 4.733 & 4.803 \\
$P(0|1)\times10^{3}$ & 7.8 & 28.3 & 15.6 & 30.3 \\
$P(1|0)\times10^{3}$ & 2.9 & 18.6 & 3.4 & 17.6 \\
Single-qubit gate error $\times10^{4}$ & 1.4 & 1.6 & 2.6 & 4.9 \\
Readout pulse duration [\textmu s] & 1.3 & 1.3 & 1.3 & 1.3 \\
ECR error $\times10^{3}$ & \multicolumn{2}{c}{3.9} &
                           \multicolumn{2}{c}{6.8} \\
ECR duration [ns] & \multicolumn{2}{c}{590} &
                     \multicolumn{2}{c}{660} \\
\bottomrule
\end{tabular}
\end{table}
In Table~\ref{tab:ibm_characterizations}, state-preparation errors are indicated as $P(\ket{1}|\ket{0})$ and $P(\ket{0}|\ket{1})$ (the probability to register $\ket{1}$ when preparing $\ket{0}$, and vice versa). Dephasing is commonly characterized by the Ramsey time $T_2^*$ and the Hahn-echo time $T_2$. Because echo sequences mitigate low-frequency noise, $T_2$ is typically longer than $T_2^*$. Our protocol leaves the sensing qubits exposed for a duration $\tau$ without echoing, so $T_2^*$ sets the practical limit on $\tau$. The Hahn-echoed $T_2$ values reported in the table correspond to IBM's device calibration measurements, while $T_2^*$ was measured by us by sampling quantum circuits containing SX gates and controlled delays. Since these values for $T_2$ and $T_2^*$ do not account for fluctuations over time (they are not averaged over multiple experiments), we conservatively set $\tau=10$ \textmu s for the test on hardware.\\
On the target backends, the only parametrized single-qubit rotation is $\text{R}_\text{Z}$, so each $\text{R}_\text{Y}(\theta/m)$ Trotter step transpiles to the sequence RZ-SX-RZ-SX. To limit the accumulation of single-qubit error, we set $m=5$.

\begin{figure}
    \centering
    \includegraphics[width=0.8\linewidth]{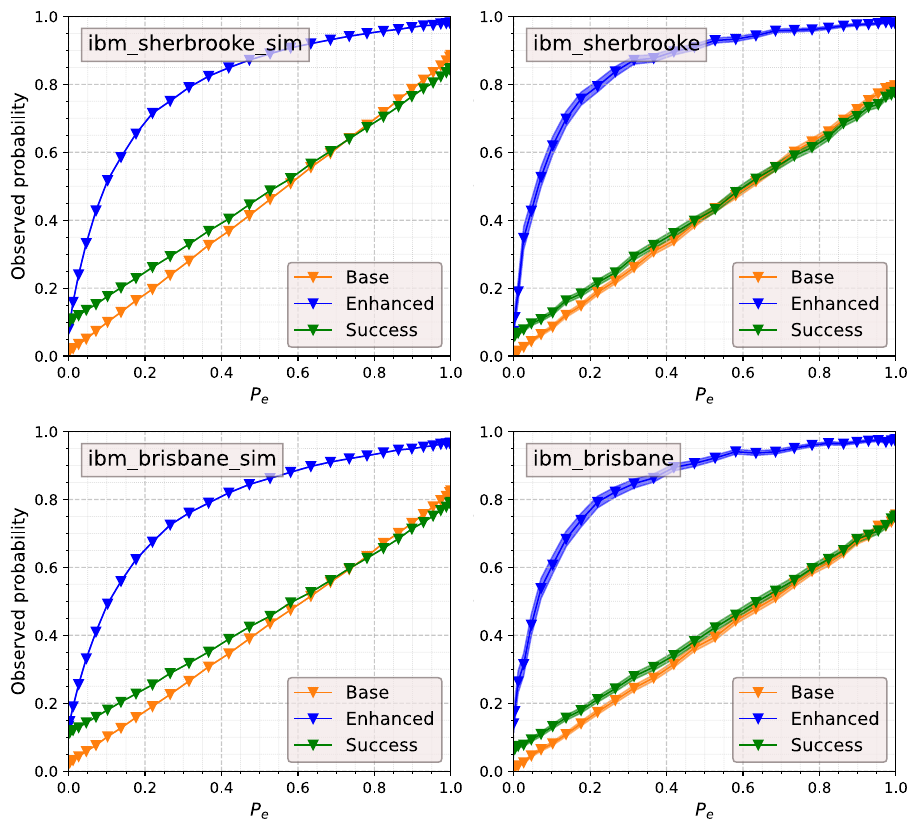}
    \caption{Comparison between noise simulation and run on IBM quantum hardware \texttt{ibm\_sherbrooke} and \texttt{ibm\_brisbane}. The blue, green, and orange curves represents $\tilde{P}^\text{obs}$, $P^\text{obs}_s$, and $P^\text{obs}$ as a function of $P_e$.}
    \label{fig:real_vs_sim}
\end{figure}

Figure~\ref{fig:real_vs_sim} compares hardware measurements with noise-model simulations for $\beta=0.4$. As a function of the ideal signal strength $P_e$, we retrieve the observed transformed probability $\tilde{P}^\text{obs}$, the protocol success probability $P_\text{success}^\text{obs}$, and the baseline, not enhanced $P_e$ sampling experiment yielding $P_\text{base}^\text{obs}$. The error bands in the plots indicate two standard deviations accounting for the Poissonian statistics. Simulations tend to be slightly more pessimistic near $P_e=0$, overestimating $P_\text{success}$, i.e.\ registering more false protocol successes, consistent with the conservative choices for assignment and preparation errors we adopted. A discrepancy could also be explained by the lack of information on the real qubits' effective temperature. In the simulations, we arbitrarily set $35$ mK.

Accurate noise modeling of complex quantum processors such as Eagle r3 is challenging and beyond the scope of this work: device parameters (coherence times, qubit frequencies, gate and readout errors) vary with time, and platform-specific effects (crosstalk, calibration drift, classical-control latency) are difficult to capture exhaustively. Nevertheless, the comparisons presented above demonstrate that the enhancement protocol can be tested directly on cloud QPUs even without pulse-level access, and CPTP-based simulations are suitable tools to carry out preliminary sensitivity analysis, such as the one conducted in the manuscript. In principle, both the baseline and the enhanced dark-matter detection experiments can be executed on existing devices.

\begin{figure}[ht]
    \centering
    \includegraphics[width=0.75\linewidth]{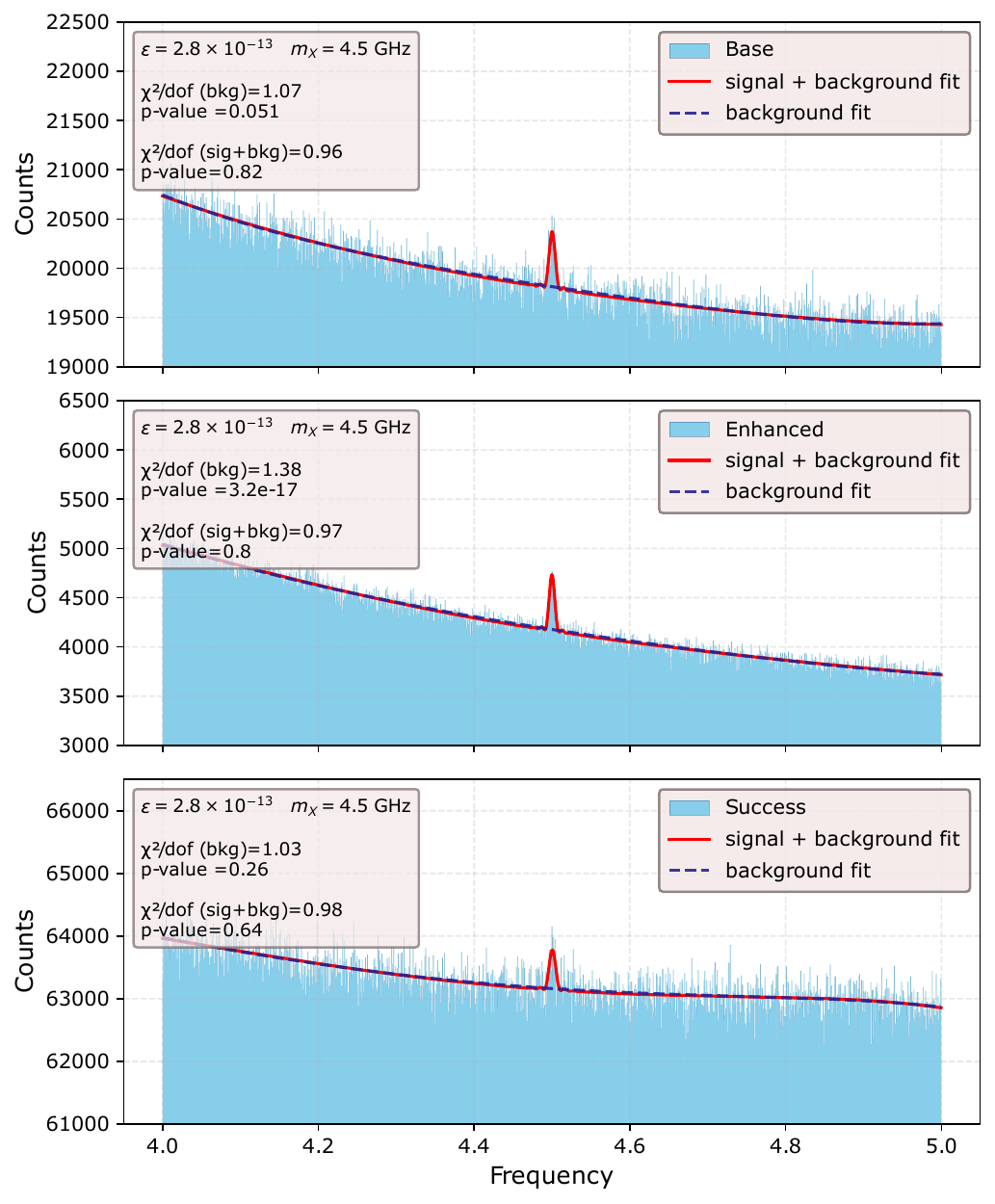}
    \caption{Example of a simulated experiment with a probe frequency sweeping between $4$ GHz and $5$ GHz with step $0.8$ MHz, setting $\epsilon=2.8\times10^{-13}$ and $m_X=4.5$ GHz. Dashed blue lines represents the null hypothesis $(H_0)$ fit, while solid red lines represent the signal $(H_1)$ fit. The number of samples per bin in the base experiment was set to $2\times 10^6$, while for the enhanced experiment $1\times 10^6$. The factor of two reflects that the enhanced protocol uses two qubits. Accordingly, we compare two independent single-qubit baseline experiments with one enhanced two-qubit experiment.}
    \label{fig:fit_analysis}
\end{figure}

\section{Statistical analysis and sensitivity extraction}
We construct synthetic datasets by evaluating the noise model across the probe-frequency and signal-strength $(f_p, P_e)$ grid, producing observed probabilities $P_{\rm base}^{\rm obs}$,  $\tilde P^{\rm obs}$, $P_s^{\rm obs}$. For each assumed $\epsilon$ (hence $P_e$), we generate counts per frequency bin and perform a binned-likelihood hypothesis test between $H_0$ (background only) and $H_1$ (signal plus background). Under a Poisson model for each frequency bin, the log-likelihood is
\begin{equation}
    \ln\mathcal{L}(\boldsymbol{\theta})=\sum_{i=1}^N \left[n_i\ln\mu_i(\boldsymbol{\theta}) - \mu_i(\boldsymbol{\theta})\right],
\end{equation}
where $\mu_i(\boldsymbol{\theta})$ is the expected count in bin $i$ given model parameters $\boldsymbol{\theta}$, $n_i$ are the observed counts, and $N$ is the total number of bins. The test statistic is the profiled log-likelihood ratio
\begin{equation}
    q = -2\big[\ln\mathcal{L}(\hat{\boldsymbol{\theta}}_0) - \ln\mathcal{L}(\hat{\boldsymbol{\theta}}_1)\big],
\end{equation}
with $\hat{\boldsymbol{\theta}}_1\in \boldsymbol{\Theta}_1$ and $\hat{\boldsymbol{\theta}}_2\in \boldsymbol{\Theta}_2$ the parameter values that maximize the likelihoods. By Wilks' theorem, $q$ asymptotically follows a $\chi^2$ distribution with $\mathrm{dof}=\dim(\boldsymbol{\Theta}_1)-\dim(\boldsymbol{\Theta}_0)$. We adopt a significance level $\alpha=0.05$ to extract 95\% C.L. exclusion limits on $\epsilon$.\\
In practice, we fit each histogram under $H_0$ with a fifth-order polynomial background,
\begin{equation}
    \lambda(x;\boldsymbol{\theta}_0) = a + b x + c x^2 + d x^3 + e x^4 + f x^5,
\end{equation}
and under $H_1$ with the same polynomial plus a signal term:
\begin{equation}
    \lambda(x;\boldsymbol{\theta}_1) = A\,\frac{\sin\big[\Omega(x-m_X)\big]}{[\Omega(x-m_X)]^2} + \lambda(x;\boldsymbol{\theta}_0),
    \label{eq:sigplusbkg}
\end{equation}
where \(\boldsymbol{\theta}_1=(A,\Omega,m_X,a,b,c,d,e,f)\). The variable $x$ is the bin center coordinate associated with each probed frequency.
In Eq. \ref{eq:sigplusbkg}, the signal term corresponds to the qubit's Rabi dynamics when driven with fixed strength $\Omega$ for a duration $\tau$, varying the detuning $m_X-x$ \cite{Nielsen2012}.
Fig.~\ref{fig:fit_analysis} shows an example fit of the base experiment and enhanced histograms (together with the protocol success count histogram) under $H_0$ and $H_1$. The reduced $\chi^2$ and associated $p$-value illustrate the improved discriminating power of the enhanced experiment.

\subsection{Including ancilla-success information}
In the enhanced protocol, we observe that the ancilla measurement carries information on $\epsilon$ also when the protocol fails (i.e.\ the ancilla collapses in $\ket{0_\text{A}}$), because $P^\mathrm{obs}_s$ depends on $P_e$. The sensing qubit histogram is conditional on the protocol's success, as we only measure the sensing qubit when the ancilla collapses in $\ket{1_\text{A}}$. Consequently, the joint likelihood factorizes into an ancilla component $\mathcal{L}_\text{A}$ and a conditional sensing component $\mathcal{L}_\text{S}$.
\begin{equation}
\begin{aligned}
\mathcal{L}(H_0) =\mathcal{L}_\text{S}(H_0|A=\ket{1})\cdot \mathcal{L}_\text{A}(H_0),\\
\mathcal{L}(H_1) = \mathcal{L}_\text{S}(H_1|A=\ket{1})\cdot \mathcal{L}_\text{A}(H_1),
\end{aligned}
\end{equation}
so that the statistical test becomes
\begin{equation}
    q \rightarrow q = -2\left[\ln \mathcal{L}(\boldsymbol{\theta}_0) + \ln \mathcal{L}^\text{s}(\boldsymbol{\theta}_0) - \ln \mathcal{L}(\boldsymbol{\theta}_1) - \ln \mathcal{L}^\text{s}(\boldsymbol{\theta}_1)\right],
\end{equation}
leveraging failed enhancement attempts that would otherwise be lost.\\

\section{Mass and $\mathbf{\beta}$-dependent exclusion limits}
\begin{figure}
    \centering
    \includegraphics[width=0.8\linewidth]{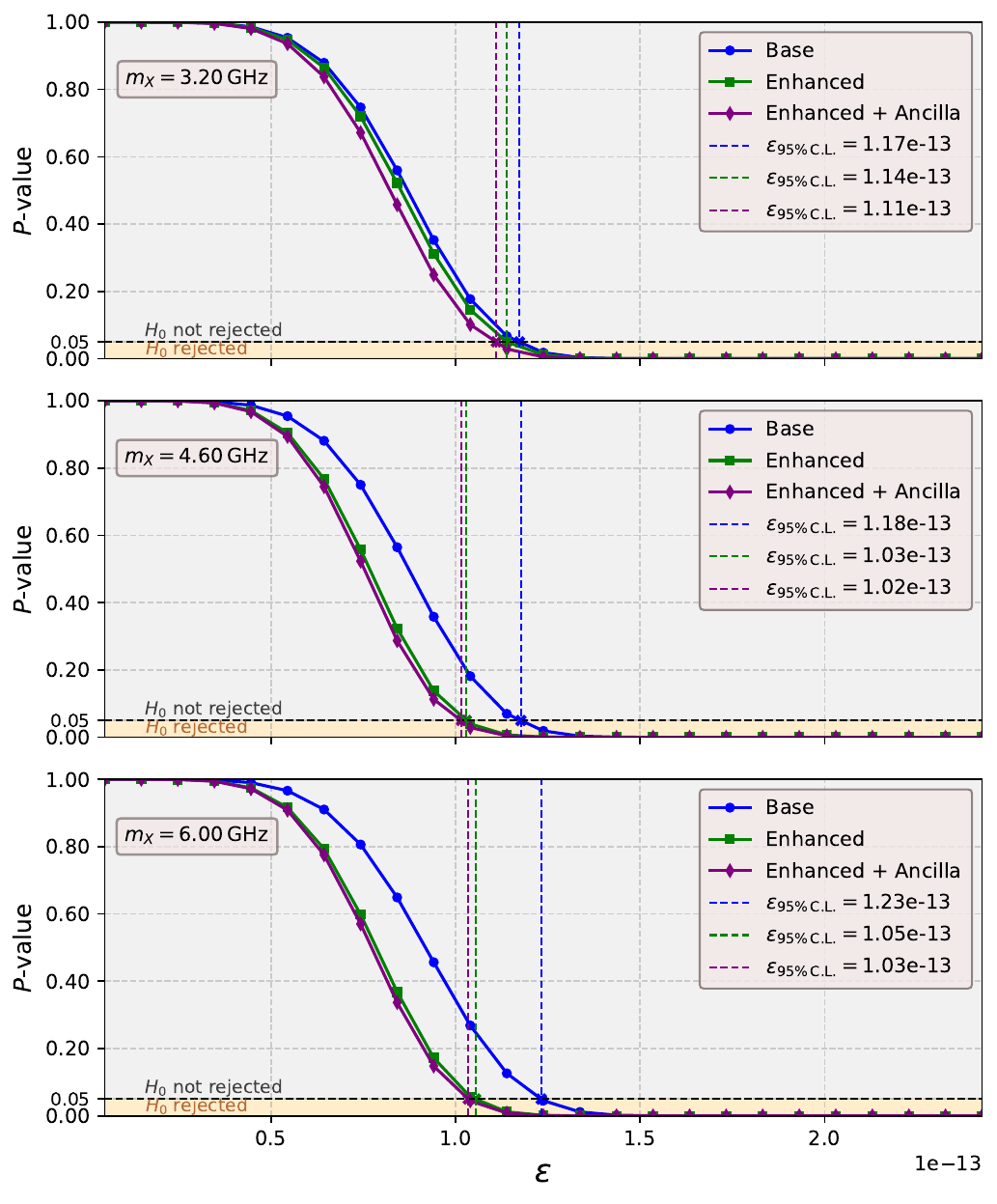}
    \caption{$p$-values of the log-likelihood ratio test statistic as a function of the kinetic mixing strength $\epsilon$, under the null hypothesis $(H_0)$ and the signal hypothesis $H_1$ (signal with strength $\epsilon$). The experiment ran with $2\times 10^6$ (base) and $1\times 10^6$ (enhanced) measurements per probed frequency. Results are shown for three dark-photon masses: $m_X = 3.2$ GHz, $m_X = 4.6$ GHz, and $m_X = 6.0$ GHz.}
    \label{fig:sensitivity_study_mass_eps}
\end{figure}

We characterized the hidden photon mass $m_X$ dependence of the exclusion limits on $\epsilon$ at $95\%$ C.L. Fig.~\ref{fig:sensitivity_study_mass_eps} depicts the $p$-values associated with the log-likelihood ratio test statistics in three cases: the baseline experiment, the enhanced experiment without additional ancilla information, and the enhanced experiment combined with ancilla information.\\
Additionally, we identified three noise scenarios and computed the $95\%$ C.L. exclusion limits on $\epsilon$ for the baseline and enhanced experiments, setting $\tau = 100$ \textmu s, $T_\text{eff}=0.35$ mK, and sweeping over $m_X$ from $2.5$ GHz to $6.0$ GHz. The noise scenarios are the following:
\begin{itemize}
    \item Medium coherence regime ($Q=\pi\times 10^6$ and $T_\phi=200$ \textmu s for all qubits), readout error rate $r=0.005$ much lower than the state preparation error $p=0.01$.
    \item Medium coherence regime ($Q=\pi\times 10^6$ and $T_\phi=200$ \textmu s for all qubits), readout error rate $r=0.05$ much higher than the state preparation error $p=0.001$.
    \item High coherence regime ($Q=2\pi\times 10^6$ and $T_\phi=400$ \textmu s for all qubits), optimal readout error rate $r=0.005$ and state preparation error $p=0.001$.
\end{itemize}
The curves confirm the larger advantage relative to the base experiment when probing higher masses than lower masses. The speedup $\mathcal{G}$ was computed at different values of the enhancement protocol rotation angle $\beta$, yielding the best results at $\beta=0.2$.
\begin{figure}[ht]
    \centering
    \begin{subfigure}{0.49\textwidth}
        \centering
        \includegraphics[width=\textwidth]{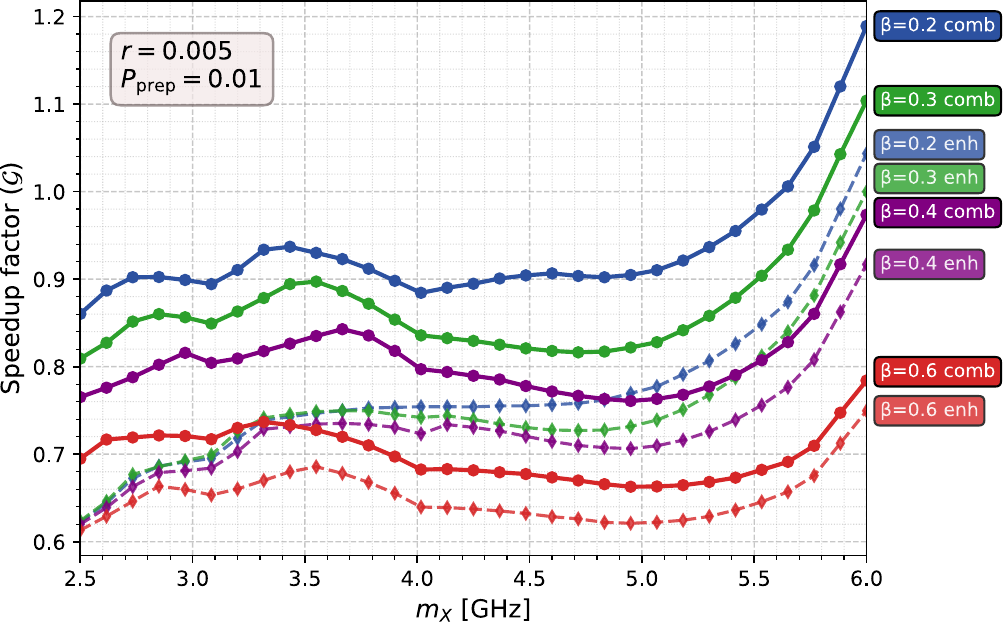}
        \caption{$r=0.005$, $p=0.01$, $Q = \pi\times 10^6$, $T_\phi=200$ \textmu s}
        \label{fig:eps_mass_bad_prep}
    \end{subfigure}
    \begin{subfigure}{0.49\textwidth}
        \centering
        \includegraphics[width=\textwidth]{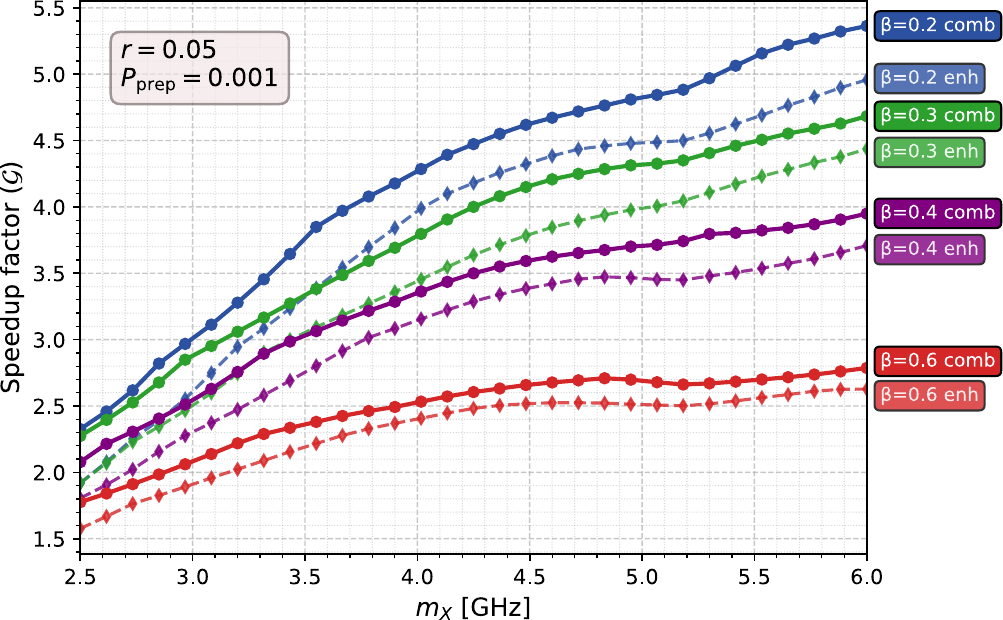}
        \caption{$r=0.05$, $p=0.001$, $Q = \pi\times 10^6$, $T_\phi=200$ \textmu s}
        \label{fig:eps_mass_bad_readout}
    \end{subfigure}
    \begin{subfigure}{0.49\textwidth}
        \centering
        \includegraphics[width=\textwidth]{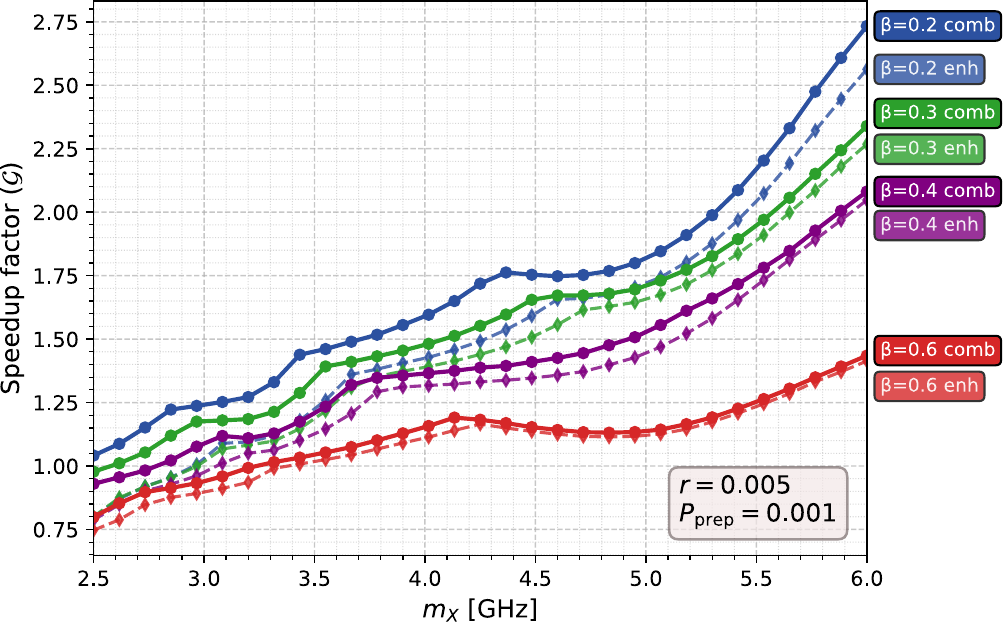}
        \caption{$r=0.005$, $p=0.001$, $Q = 2\pi\times 10^6$, $T_\phi=400$ \textmu s}
        \label{fig:eps_mass_optimistic}
    \end{subfigure}
    \caption{Hidden photon mass-dependent speedup factor of the two-qubit enhanced experiment over two single-qubit baseline experiments in parallel. Label \emph{enh} refers to the exclusion limits on $\epsilon$ extracted without additional ancilla information, while label \emph{comb} refers to the same experiment combining the ancilla-extracted success rate probability of the protocol scheme.}
    \label{fig:eps_mass}
\end{figure}

\bibliography{refs}